\documentclass{raa}           

\usepackage{graphicx,times}\usepackage{natbib} 
\usepackage{amssymb,amsmath}
\bibpunct{(}{)}{;}{a}{}{,}

\usepackage[a4paper=true,pagebackref=true]{hyperref}
\hypersetup{colorlinks = true, linkcolor = green, anchorcolor = red, citecolor = blue, filecolor = red, pagecolor = red, urlcolor = red}

\begin{document}

	\title{Has LIGO Detected Primordial Black Hole Dark Matter ?}
	\subtitle{ - Tidal Disruption in Binary Black Hole Formation}
	
    \volnopage{Vol.0 (20xx) No.0, 000--000}      
    \setcounter{page}{1}          
    
    \author{Yuan Gao \inst{1}
    \and Xiaojia Zhang \inst{2}
    \and Meng Su \inst{3}}
    
    \institute{Department of Physics, The University of Hong Kong,
Hong Kong SAR, China; {\it u3502132@hku.hk}\\
	\and Department of Earth Sciences, The University of Hong Kong, Hong Kong SAR, China; {\it xzhang17@hku.hk}\\	
	\and Department of Physics and Laboratory for Space Research, The University of Hong Kong,
Hong Kong SAR, China; {\it mengsu84@hku.hk}\\	
\vs\no
    {\small Received~~20xx month day; accepted~~20xx~~month day}}
    
	\abstract{ The frequent detection of binary mergers of $\sim 30 M_{\odot}$ black holes (BHs) by the Laser Interferometer Gravitational-Wave Observatory (LIGO) rekindled researchers' interest in primordial BHs (PBHs) being dark matter (DM). In this work, we looked at PBHs distributed as DM with a monochromatic mass of $30 M_{\odot}$ and examined the encounter-capture scenario of binary formation, where the densest central region of DM halo dominates. Thus, we paid special attention to the tidal effect by the supermassive black hole (SMBH) present. In doing so, we discovered a necessary tool called loss zone that complements the usage of loss cone. We found that the tidal effect is not prominent in affecting binary formation, which also turned out insufficient in explaining the totality of LIGO's event rate estimation, especially due to a microlensing event constraining the DM fraction in PBH at the mass of interest from near unity to an order smaller. Meanwhile, early-universe binary formation scenario proves so prevailing that the LIGO signal in turn constrains the PBH fraction below one percent. Thus, people should put more faith in alternative PBH windows and other DM candidates.
	}

   \titlerunning{Has LIGO Detected PBH DM}
  
	\maketitle
\section{Introduction}
\label{sect:intro}
The advent of the gravitational wave (GW) astronomy has opened a new gate in probing the universe. Since the first confirmed binary BH merger event in 2015, there have been nine more detections of the same type and one of binary neutron  stars (NSs) \citep{2016PhRvL.116f1102A,2016PhRvL.116x1103A,2017PhRvL.118v1101A,2017ApJ...851L..35A,2017PhRvL.119n1101A,2019PhRvX...9c1040A}. All binary BH events feature component mass on the order of $\sim$ 10 solar masses ($M_{\odot}$) or beyond, and 7 reach the order of $30 M_{\odot}$ or above. 

On one hand, such masses are generally too heavy for stellar BHs formed from dying stars, whose mass is usually on the order of $\sim 1 M_{\odot}$. A more likely candidate might be PBHs that formed from gravitationally collapsed over-densities in the radiation dominated era. The formation of PBHs is not related to stellar gravitational collapse, and is majorly attributed to an increased cosmological energy density, thus their masses could span a much wider range (\citealt{2016PhRvD..94h3504C}). 

On the other hand, the frequent detection suggests a relatively abundant underlying population, which reminds people of the possibility that PBHs might constitute DM. DM is a hypothetical model motivated by observations such as the flattening of galaxy rotation curves (see, for example, \citealt{2000MNRAS.311..441C}) that implies the existence of a considerable amount of unknown masses throughout the galactic region. It is now believed to comprise $\sim 85 \%$ of all matter content of the universe. As PBHs formed in the radiation-dominated era, they are essentially non-baryonic and well qualify for constituting DM (\citealt{2016PhRvD..94h3504C}), an idea arose since the earliest studies of PBHs (\citealt{1975Natur.253..251C}). And despite various observational constraints on the fraction of DM in PBHs accros their wide range of masses, $30 M_{\odot}$ hasn't been ruled out for PBH to constite a major part of DM (\citealt{2015PhRvD..92b3524C,2016PhRvL.116t1301B}). Non-detection in 30 years of searches for an alternative DM candidate called weakly interacting massive particles (WIMPs) such as supersymmetry particle or the axion (\citealt{1996PhR...267..195J, PRESKILL1983127}) also encouraged the study of PBHs' candidature. 

Thus it is intriguing to ask if LIGO has detected PBH DM, and whether its GW detection could shed lights on the the mass, fraction etc. of PBHs in constituting DM. Specifically, we would like to know whether PBHs distributed as DM would form binary mergers with an event rate compatible with LIGO's estimation of $9.7-101 Gpc^{-3} yr^{-1}$ (\citealt{2019PhRvX...9c1040A}). To do so, we first decided on the appropriate population distribution of the proposed PBH DM, then looked at the condition of binary PBH formation. As a starting point, we focused on PBHs with a monochromatic mass function of $\sim 30 M_{\odot}$ in a Milky Way (MW) like galaxy. And throughout this study, we worked under the unit system with $G=c=1$. 

To find out the PBH DM distribution, we noted that N-body simulations of the dark halo have shown the formation of a dense core at the center distributed as a power law with index $\sim -1$, though a precise parametric profile has not been resolved between several candidates such as the Navarro–Frenk–White (NFW) and Einasto (\citealt{profiles}). However, this only takes into account the particle DM component without involving the stars and the SMBH at the galaxy center. At their presence, the density slope in the central region is believed to be further steepened to a power of $\sim -2$ (\citealt{massseg}) due to segregation of the PBHs, yet should vanish at the capture radius of the SMBH (\citealt{spike}), resulting in a 'spike' density profile. Thus we considered all mentioned effects, and adopted the most appropriate distribution within their validity ranges. 

PBHs can form binaries via two-body and three-body encounters. \cite{1989ApJ...343..725Q, 1993ApJ...418..147L} have shown that binary formation via three-body encounter is only non-negligible in small halos and generically leads to wide binaries that won't merge within the Hubble time. Thus in this work we focused on the two-body scenario where two PBHs emit enough GW energy during a close encounter and capture each other. In such case the galaxy center where DM is the densest shall dominate binary encounter, and we calculate the capture rate using current best knowledge of the dark halo distribution. We also take into account the SMBH's tidal effect, which is most prominent at the galaxy center as well. Then by convolving the rate calculation in a single MW-like halo with the halo mass function, the merger rate density per comoving volume can be compared with LIGO's estimation of $9.7 \sim 101 \; Gpc^{-3}\,yr^{-1}$. This might side-support or constrain the PBH DM proposal.

The rest of this paper is arranged as follows. We review the available density distributions that combine the properties of both DM and PBH, and decide on the most appropriate one for this work in section \ref{sect:distr}. Then we look at binary capture mechanism in section \ref{sect:capt}, and find the effective binary formation rate in section \ref{sect:tid} taking into account the tidal disruption effect. We present and discuss the results in section \ref{sect:result}, and draw a conclusion in section \ref{sect:concl}.  

\section{Dark Matter Distribution in the Primordial Black Hole Perspective}
\label{sect:distr}
The NFW profile is by far a widely accepted dark matter halo density distribution, obtained by N-body simulation of resolved dark matter particle evolving from some initial density perturbation spectrum (\citealt{NFW}). The mass density distribution is described as: 
\begin{equation}
\frac{\rho(\vec{r})}{\rho_{crit}}=\frac{\delta_c}{r_*(1+r_*)^2}
\end{equation}
where $\rho_{crit}$ is the critical density of the universe, $r_*$ is the galactocentric distance normalized by a scale radius $R_s$ which relates to the virial radius by a concentration parameter $c=\frac{R_{vir}}{R_s}$ that is dependent on the halo mass. 

This is not accurate, however, as it only takes account of the dark matter particles without considering the baryonic component of the galaxy. It becomes most important especially near the galaxy center, where a SMBH is generally believed to reside, which steepens the potential well. Here we compare two models that take care of such effect: mass segregation and DM spike. However, these studies are based on the MW, and we start by examining the case of our own galaxy first.

The effect of mass segregation states that, according to simulation, for a mixed population with lighter but dominant stars and heavier but non-dominant stellar objects - for which PBHs qualify, the existence of the SMBH would result in the heavier population segregating into the center, forming a cusp with power index $\sim -2$ (\citealt{massseg}). Although baryonic dominance in the inner galaxy hasn't been universally validated, studies have provided some confirmation for disk galaxies, summarized in \cite{bmdom}. Here as we first look at the inner region of the MW, baryonic dominance is highly probable and the mass segregation model is applicable. Moreover, we also found the result in agreement with the spike model to be introduced later, which adds to our confidence. According to \cite{massseg}, various studies have agreed that $\sim$ 20,000 10$M_\odot$ BHs should have segregated into the inner 1pc of a MW like galaxy, by which we normalize the mass distribution of the segregated profile. 
\begin{equation}
\rho_{seg}=\frac{7.7\times10^{-10}}{r^2} m^{-2}
\label{segdens}
\end{equation}

The drawback of this model is that it does not go deep into the galaxy center where the density starts to drop and vanishes by the capture radius of the SMBH $r_{cap}=4R_\bullet$, 4 times its Schwarzschild radius  (\citealt{caprad}). Within this distance, objects are believed to directly plunge into the SMBH. Thus we considered a second approach, which studied the effect of SMBH on the paprticle DM distribution like NFW, and also agrees with a $\sim-2$ cusp but does vanish properly (\citealt{spike}). The profile is called a 'spike' to distinguish from the segragated cusp. This study, however, does not consider the gravitational effect due to the stellar component, and is restricted to the region of $\leq 0.2$ pc where the SMBH dominates the gravitational potential. Thus we could treat this as a more accurate description of the PBH DM distribution at the innermost galaxy center. From an initial distribution of $\rho(r)_i = \rho_0 (r/r_0)^{-\gamma}$, $0 < \gamma < 2$, the spike density profile is given by:

\begin{equation}
\rho(r)_{sp} = \rho_0\left(\frac{R_{sp}}{r_0}\right)^{-\gamma} \left(1-\frac{r_{cap}}{r}\right)^3 \left(\frac{R_{sp}}{r}\right)^{\gamma_{sp}} 
\end{equation}
where $ \gamma_{sp} = \frac{9-2\gamma}{4-\gamma} $, $ R_{sp} = \alpha_{\gamma} r_0 \left(\frac{M_\bullet}{\rho_0 r_0^3}\right)^{\frac{1}{3-\gamma}}$, and $M_\bullet$ is the mass of the SMBH. $\alpha_{\gamma}$ is numerically derived for different values of $\gamma$, where for an NFW initial profile with $\gamma=1$, it was taken to be 0.122. Concretely, in the case of MW:

\begin{equation}
\rho(r)_{sp} = \frac{1.04 \times 10^{-4}}{r^\frac{7}{3}} \left(1-\frac{5.04 \times 10^{10}}{r}\right)^3  \;m^{-2}
\label{spike}
\end{equation}  

\begin{figure}
 \centering
 \includegraphics[width=100mm]{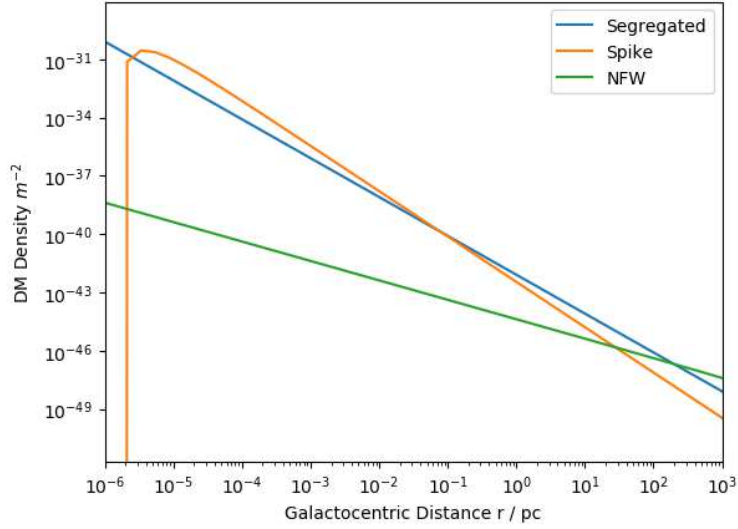}
 \caption{DM density distributions}
\label{fig:profile}
\end{figure}

The three profiles (NFW, segregated and spike) for a MW-like galaxy are plotted in Figure~\ref{fig:profile} for comparison.

It turned out that different profiles take charge of their validity ranges just fine. The spike profile is most accurate within $\sim 0.2 \;pc$ where SMBH dominates stellar dynamics, and it hands over to the segregated profile where stellar objects play a large part, and is finally replaced by the NFW profile at $\sim 200 \;pc$ where DM starts to dominate the population and its N-body simulation becomes more accurate. In the following calculation of binary capture rate, we will make use of this composite of density distributions.   

\section{Capture of Binary Black Holes}
\label{sect:capt}

A BH binary can form if ample energy is lost via GW during a close encounter. The situation is specified by 4 parameters, $m_1, m_2, \vec{r_r}, \vec{w}$, namely the masses, relative position and relative velocity. 

The first step is to consider the energy loss along a complete Keplerian hyperbolic orbit, assuming 2 BHs approaching each other from infinity. As energy radiation along the path drives the binary closer than the unperturbed orbit, more energy will be actually lost. Thus this approximation provides a conservative maximum impact parameter $b$ for capture. In this case, the magnitude of the relative velocity $w$ and impact parameter $b$ will take the place of the 2 vector parameters. \cite{Turner77} calculated the energy loss for an arbitrary unbound orbit at the Newtonian limit, which is the most reliable result used today. Though the non-relativistic approximation is obvious, an analytic solution taking care of the strong gravitaional field near periapsis has not yet been obtained. We will stick to Turner's result here.

Setting $G=c=1$, the radiation is given in \cite{Turner77} by 
\begin{equation}
\delta E=\frac{8}{15}M\eta^2w^7f(e)
\end{equation} 
where M is the total mass of the bianry, $\eta$ is the symmetric mass ratio $\frac{m_1m_2}{M^2}$, $e$ is the orbital eccentricity given by $\sqrt{1+\frac{b^2w^4}{M^2}}$, and $f(e)$ is an enhancement factor given by
\begin{equation}
f(e)=\frac{24\cos^{-1}(\frac{1}{e})(1+\frac{73}{24}e^2+\frac{37}{96}e^4)+\sqrt{e^2-1}(\frac{301}{6}+\frac{673}{12}e^2)}{(e^2-1)^{\frac{7}{2}}}
\end{equation}

\begin{figure}
\centering
 \includegraphics[width=100mm]{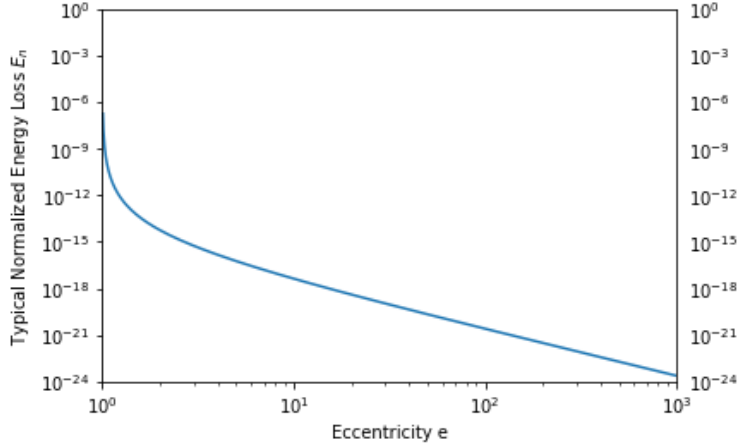}
 \caption{Normalized energy loss with typical relative velocity $\sim 150 km/s$ at equal-mass binary setup, over orbital eccentricity }
\label{fig:fe}
\end{figure}

For binary capture, $\delta E$ should exceed the kinetic energy $\frac{\eta Mw^2}{2}$. For a Milky-Way like galaxy of virial velocity  $\sim 150 km/s$, energy loss normalised by kinetic energy equals
\begin{equation}
E_n \equiv \frac{\delta E}{KE}=\frac{16}{15}\eta w^5f(e)\leq\frac{4\times 10^{-15}}{15}f(e)
\end{equation} 
where equality holds when two masses are equal. This is plotted in Figure~ \ref{fig:fe}. From the plot we can observe that the capture condition can only be met with $e \sim 1$. Thus we can restrict our attention to near-parabolic orbits only. 

At this limit the maximum impact parameter is given by 
\begin{equation}
b_{max}=M[\frac{340\eta}{3cos(1)w^9}]^\frac{1}{7}
\label{eqn:bmax}
\end{equation} 
Using this, we could verify that $e$ is indeed very close to 1. From $b_{max}$ the capture cross section is given by
\begin{equation}
\sigma=\pi b_{max}^2
\label{eqn:sigma}
\end{equation}
And without considering tidal effect, a naive capture rate can be calculated as 
\begin{equation}
R_1=\int\frac{n(\vec{r},\vec{v_1})n(\vec{r},\vec{v_2})}{2}\sigma w d\vec{r}d\vec{v_1}d\vec{v_2}
\label{eqn:naiverate}
\end{equation} where n is the PBH number density. From this very simple consideration, we see that the capture rate is independent of the PBH mass, as $n^2 \sim m_{PBH}^{-2}$, and $\sigma \sim m_{PBH}^2$ as viewed from equation (\ref{eqn:bmax}) and equation (\ref{eqn:sigma})

\section{Tidal Disruption}
\label{sect:tid}
From section \ref{sect:distr} we have seen that PBH DM density peaks near the SMBH, where the tidal effect is significant. This would influence binary formation in at least 2 ways: on one hand the capture condition must also incorporate the tidal force, that the lost energy in GW should result in a binary close enough to resist the tidal tear; on the other hand, if some formed binaries move too close to the SMBH, the tidal force might become strong enough to break the binary. It is evident that from the first effect, not the whole capture cross section is effective; from the second effect, we need to determine the directions of motion that will result in tidal disruption of formed binaries, known as the 'loss cone' (\citealt{losscone}), and exclude the binary population within. We will take account of these two aspects in this section. 

To quantify the tidal effect, we note that it is dependent on the object's size, in this case being the binary semi-major axis. For a binary of mass M and semi-major axis a, the tidal force outweighs its gravitational attraction if the binary falls closer to the SMBH than the tidal distance (\citealt{tidalr})

\begin{equation}
r_t=a \sqrt[3]{\frac{M_\bullet}{M}}
\label{eqn:rt}
\end{equation} 
The actual disruption distance would be the larger of $r_t$ and $r_{cap}$, which we denote as $r_d$. 

The semi-major axis is dependent on capture conditions (\citealt{a}):

\begin{equation}
a=\frac{M}{w^2 [(\frac{b_{max}}{b})^7-1]}
\label{eqn:a}
\end{equation}
Thus it is evident that the actual impact parameter $b$ within $b_{max}$ shall be closely examined, as it results in a captured binary with different sizes, which correspond to different disruption distance $r_d$. 

We also note that the binay size is positively related to the mass of the binary, and the same goes for the tidal disruption radius. Thus heavier PBHs tend to form wider binaries which are subject to larger disruption radius. Combined with results inferred from section \ref{sect:capt}, we can expect lighter PBH binaries less affected by the tidal effect.

\subsection{Effective Binary Capture}
\label{subsect:effcap}
In section \ref{sect:capt}, the capture condition does not take into account the tidal force, thus we name it a naive capture. If such a capture happens within the disruption distance to the SMBH, the binary won't form. Thus an effective binary capture must happen outside $r_d$. We consider $30M_\odot$ PBHs comprising all DM, moving isotropically at velocity calculated differently for different regions. Within the spike profile, we take $v=\sqrt{\frac{M_\bullet}{r}}$ as DM is negligible to the mass of SMBH. Beyond that, DM mass cannot be neglected, and for simplicity we use the virial velocity of the galaxy $v_{vir}\sim 150 km/s$. Based on equation (\ref{eqn:naiverate}), the effective capture rate $R_{eff}$ is calculated by excluding those naively captured binaries within the disruption distance:

$$R=\int\int\frac{n(\vec{r},\vec{v_1})n(\vec{r},\vec{v_2})}{2} w d\vec{r}d\vec{v_1}d\vec{v_2}\int_0^{b_{max}(w)}2\pi b db \Theta(r-r_d(w,b))$$
$$=\int\frac{n(\vec{r})^2}{2}4\pi r^2 dr\int\frac{w}{(4\pi)^2}d\vec{v_1}d\vec{v_2}\int_0^{b_{max}} 2\pi b db \Theta(r-r_d)$$
$$=\int\frac{n(\vec{r})^2}{2}4\pi r^2 dr\int\frac{w}{(4\pi)^2}d\vec{v_{com}}d\vec{w}\int_0^{b_{max}}2\pi b db \Theta(r-r_d)$$
$$=\int\frac{n(\vec{r})^2}{2}4\pi r^2dr\int\frac{2vsin\frac{\theta_r}{2}}{4\pi}2\pi sin\theta_rd\theta_r\int_0^{b_{max}}2\pi b db \Theta(r-r_d)$$
\begin{equation}
=\frac{8\pi^2}{m^2} \int\rho(\vec{r})^2 r^2 v dr \int (sin\frac{\theta_r}{2})^2 cos\frac{\theta_r}{2} d\theta_r \int_0^{b_{max}}b db \Theta(r-r_d)
\label{eqn:effectiverate}
\end{equation}
where we have included a Heaviside function $\Theta(r-r_d)$ to exclude those naively captured binaries within the disruption distance, without which the integral reduces to equation (\ref{eqn:naiverate}). In the second line we used isotropy in the velocity space, whose independence between 2 populations enables us to change to center-of-mass (C.O.M.) and relative velocity spaces in line three, and by integrating out irrelevant variables in the second integral we kept the polar angle $\theta_r$ between 2 velocities. 

Note that the mass at the front does not denote dependence, as it also appear in $b_{max}$ as in equation (\ref{eqn:bmax}). Nor is the radial integral self contained, as $r$ appears in the last integral as well. The only relation is that the effective capture rate scales with density squared.

As stated before, we will use the spike profile to account for the region within 0.2 $pc$, segregated profile out to 200 pc, and NFW outwards. Results are summarized in Table~\ref{tabl}, with the naive capture rate also listed for comparison.  

\subsection{Loss Zone and Loss Cone}
\label{subsect:lzlc}

Upon effective binary formation, some fraction moving too close to the SMBH would gradually enter the disruption distance before itself merging. This range of directions was given the name loss cone (\citealt{losscone}). For a specific binary with semi-major axis a, there is a critical angular momentum $l_{crit}$ that the binary would graze the sphere of tidal disruption. 

\begin{equation}
\label{lcrit}
l_{crit}=r_d \sqrt{v_{com}^2-\frac{2M_\bullet}{r}+\frac{2M_\bullet}{r_d}}
\end{equation}

\begin{figure}
 \centering
 \includegraphics[width=80mm]{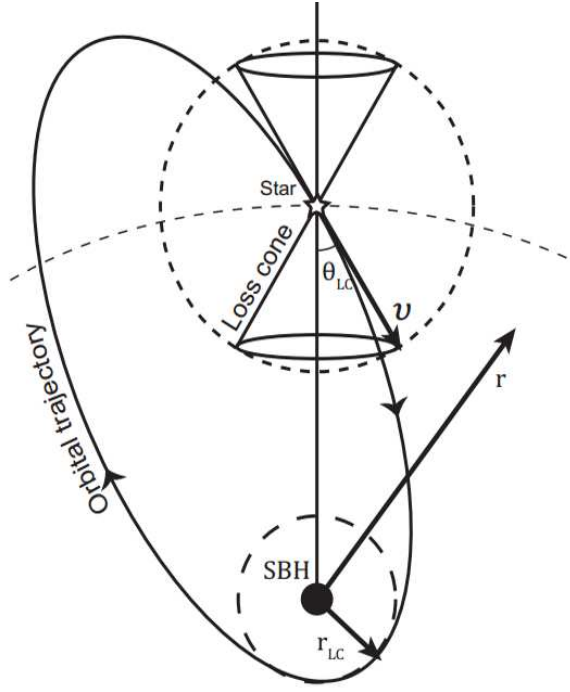}
 \caption{Illustration of loss cone, where the binary is positioned at the 'Star' position, and SBH stands for Supermassive Black Hole}
\label{fig:lc}
\end{figure}

The corresponding directions of motion form the boundary of the loss cone, illustrated in Figure~\ref{fig:lc} directly adopted from \cite{losscone}.  

Thus to account for this effect, we need to exclude those effectively captured binaries moving within the loss cone. In the literature, loss cone has been assumed small (\citealt{losscone}). We found, however, by comparing the critical angular momentum (eqn. \ref{lcrit}) with $v \times r$ for a binary with its corresponding $r_d$, there is a radial distance within which the critical angular momentum gets larger than $v\times r$, which means the binary will always enter the disruption distance whichever direction it moves; or equivalently, the loss cone covers the whole solid angle. We name this radial distance the 'loss zone' distance, for a specific binary configuration. 

\begin{equation}
r_z=r_d \frac{\sqrt{1+\frac{8M_\bullet}{v_{com}^2r_d}}-1}{2}
\end{equation}

Thus the loss cone angle should be computed as:
\begin{equation}
 \theta_l=
  \begin{cases}
  arcsin(\frac{l_{crit}}{v_{com}r}) & r>r_z, r>r_d \\
  \frac{\pi}{2} & else 
 \end{cases}
\end{equation}

To account for this effect, equation (\ref{eqn:effectiverate}) should be further modified, where the integration of binary C.O.M. directions of motion should be screened as well.

$$R=\int \frac{n(\vec{r})^2}{2} 4\pi r^2 dr
  \int \frac{w}{4\pi} d\vec{w}
  \int_0^{b_{max}} 2\pi b db 
  \int_{\theta_l}^{\pi-\theta_l} \frac{1}{4\pi} d\vec{v_{com}}$$

\begin{equation}
 =\frac{8\pi^2 \sqrt{M_\bullet}}{m^2} \int\rho^2 r^{1.5}dr \int sin\frac{\theta_r}{2}^2 cos\frac{\theta_r}{2} d\theta_r \int_0^{b_{max}}b db cos\theta_l
 \label{eqn:survivalrate}
\end{equation} 

The binary capture rates for different regions are calculated and added to Table~\ref{tabl} as well.

\section{Results and Discussion}
\label{sect:result}
We list and compare 3 binary formation rates with more physics coming into play, and also look at different regions of interest where the analytic profiles differ.

\begin{table}[h]
\centering
\caption{Capture Rate at Different Regions} 
 \begin{tabular}{| c | c | c | c |} 
 \hline
 Distance & $r_{cap}$ - 0.2 pc & 0.2 - 200 pc & 200 - Mpc\\
 \hline
 Profile & Spike & Segregated & NFW\\ 
 \hline
 Naive capture/halo & $5.60\times 10^{-8} /yr$ & $6.24 \times 10^{-11} /yr$ & $6.29 \times 10^{-9} /yr$\\
 \hline
 Effective capture/halo & $3.93\times 10^{-8} /yr$ & $5.73 \times 10^{-11} /yr$ & $5.82 \times 10^{-9} /yr$ \\
 \hline
 Outside Loss Cone/halo & $3.69 \times 10^{-8}/yr$ & $5.64 \times 10^{-11} /yr$ & $5.74 \times 10^{-9} /yr$\\
 \hline
 \end{tabular}
\label{tabl}
\end{table}

It is immediately evident that the spike region dominates binary formation, while cumulatively the outskirt NFW region contributes more than the segregation region. We could also observe that the tidal force eliminates $\sim 30 \%$ naively captured binaires in the spike, and the majority of the rest survive the following evolution without entering the loss cone. This effect dies down as we move away from the galaxy center, and overall speaking the tidal effect does not influence binary formation rate noticeably by orders of magnitude.

To further illustrate the spatial details of binary formation, we plotted the capture rate density and cumulative capture rate for the 3 different methods in Figure~\ref{fig:dens} and \ref{fig:cumu} below. 

\begin{figure}[h]
 \begin{minipage}[t]{0.495\linewidth}
 \centering
 \includegraphics[width=80mm]{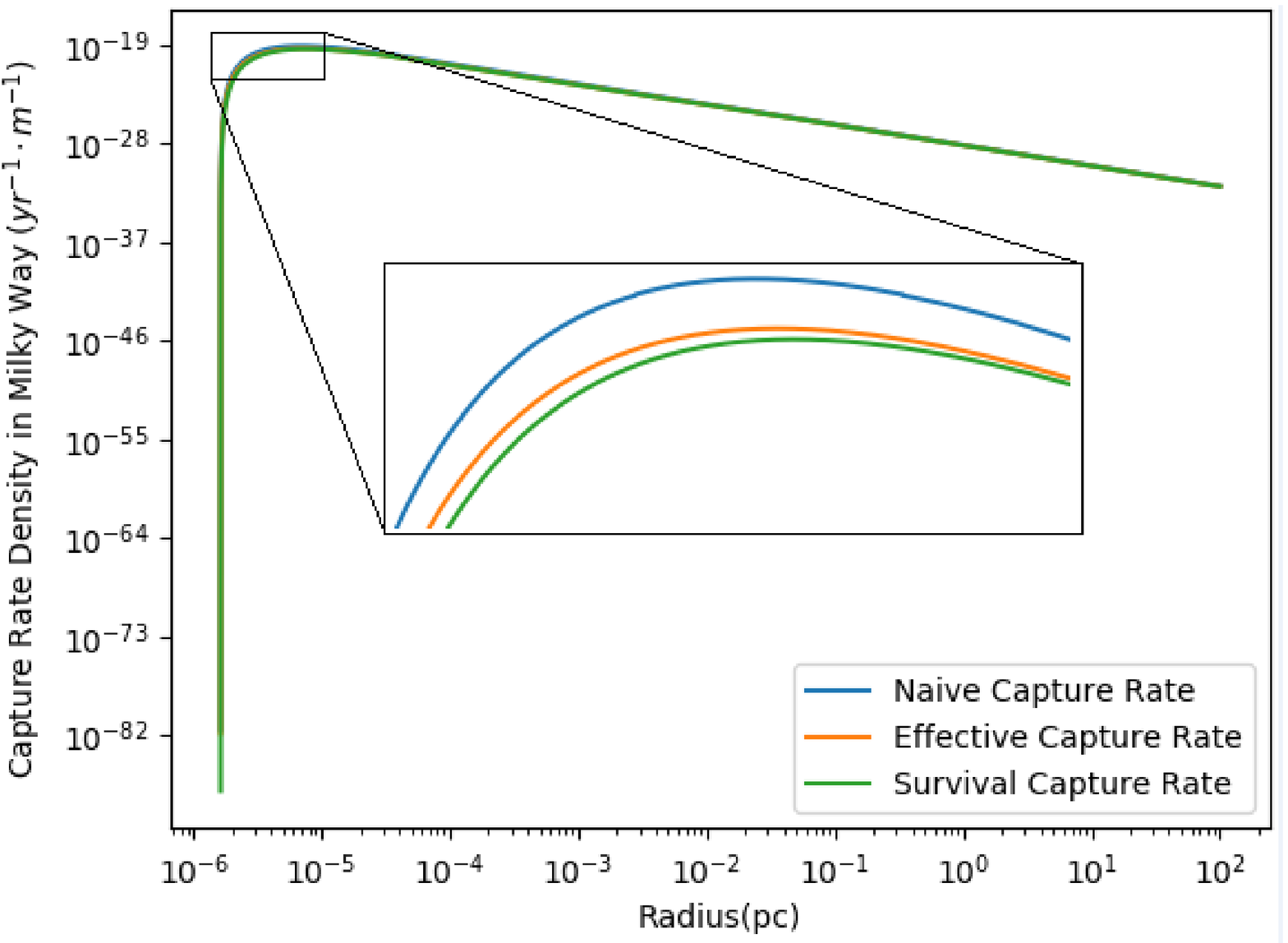}
 \caption{Capture rate density in MW, with window showing the peak region, where 3 curves differ the most.}
 \label{fig:dens}
 \end{minipage}
 \begin{minipage}[t]{0.495\textwidth}
 \centering
 \includegraphics[width=80mm]{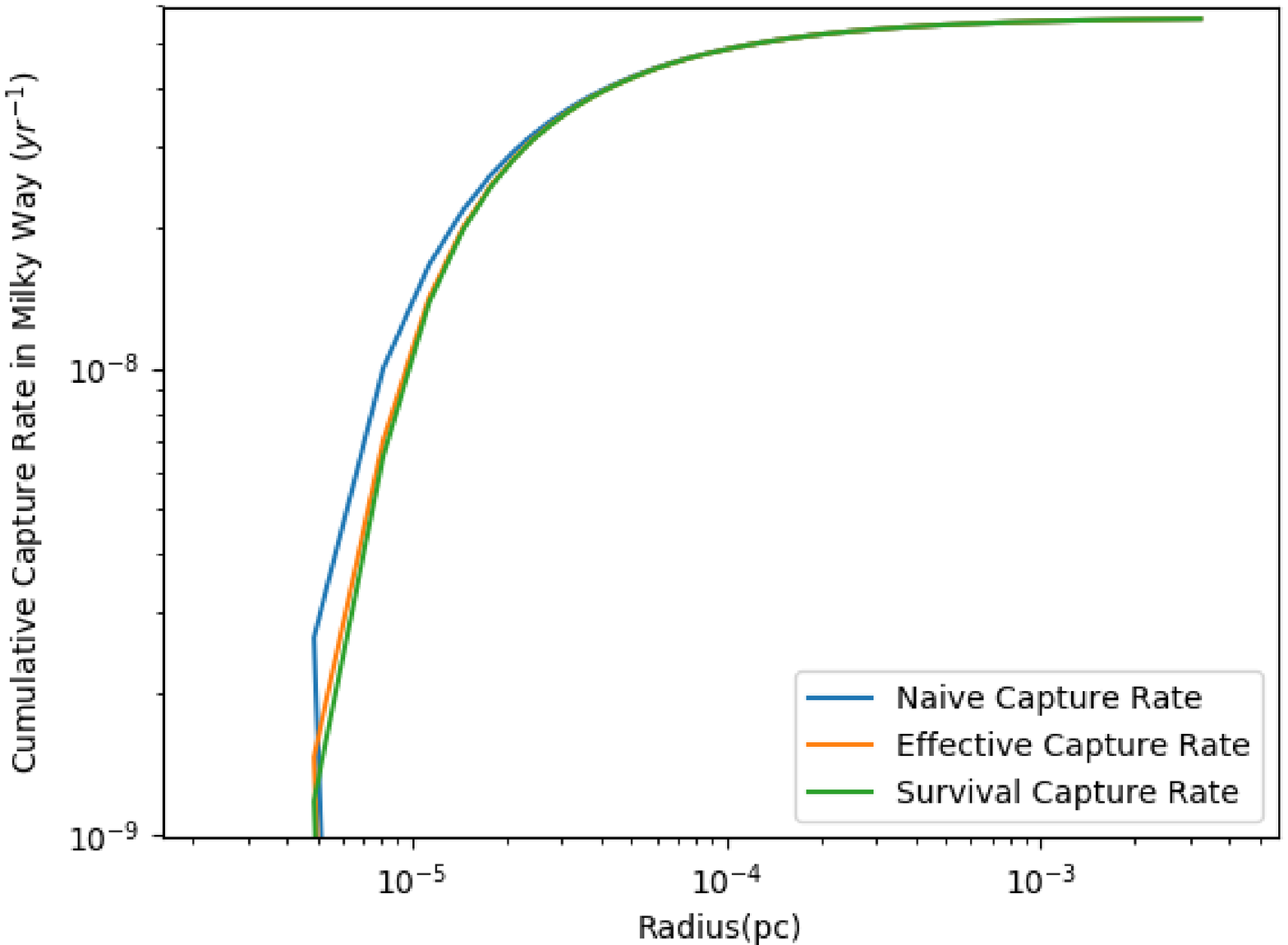}
 \caption{Cumulative capture rate in the MW}
 \label{fig:cumu}
 \end{minipage}
\end{figure}

\begin{figure}
\centering
 \includegraphics[width=\textwidth]{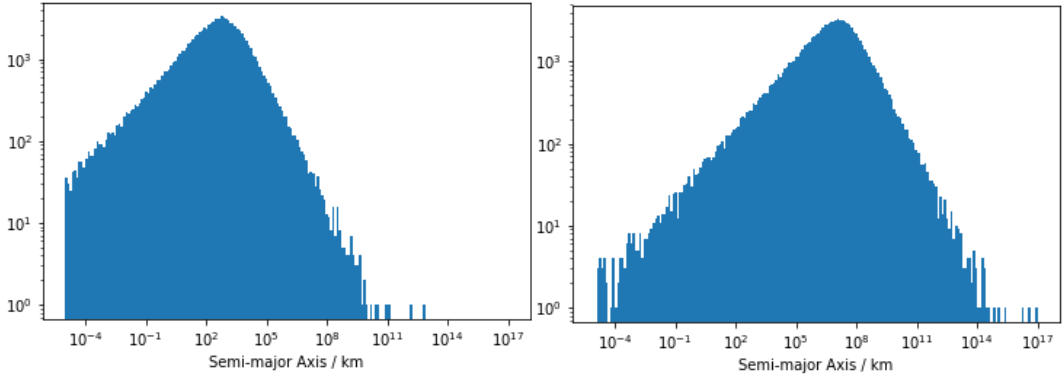}
 \caption{Counts of simulated binary size for one million runs. Left: at $10^{-5} \;pc$. Right: at $ 0.2 \;pc $ where the validity region of the spike profile was assumed to terminate}
\label{fig:simu}
\end{figure}

From the plots we see that the 2 rates taking into account the tidal effect do not differ from the naive binary capture rate noticeably, thus the SMBH's tidal disruption plays no significant role in binary formation. To find out the reason, we notice that the maximum impact parameter $b_{max}$ given by equation (\ref{eqn:bmax}) is inversely proportional to the magnitude of relative velocity by $w^{-\frac{9}{7}}$. Thus in the innermost region where the SMBH mass dominates the gravitational potential, the maximum impact parameter shrinks closer to the center, resulting in harder binaries. We also notice that the semi-major axis of captured binary, given by equation (\ref{eqn:a}), is also inversely proportional to the relative velocity by $w^{-2}$. To look at the dependence on the other factor of $(\frac{b_{max}}{b})^7-1$, we simulated a million binary formation events at 2 positions of interest to observe the distribution of binary sizes, which is shown in Figure~\ref{fig:simu}. Wen doing so, we assumed the same isotropy in relative velocity, i.e. $p(\theta_{rel})=\frac{1}{4\pi}, w=2v(r)sin\frac{\theta_{rel}}{2}$, where $v(r)$ is the velocity of PBHs being $\sqrt{\frac{M_\bullet}{r}}$ as in section \ref{subsect:effcap}. For the probability distribution of impact parameter $b$, we assumed a uniform probability within the total capture cross section: $p(b)=\frac{2\pi b}{\sigma}=\frac{2\pi b}{\pi b_{max}^2}$.

From the plot we can confirm that closer to the galaxy center, the overall binary sizes do shrink. The reduction in capture cross section and increase in PBH velocity results in the compactness of captured binaries. On the other hand, the reduction in capture cross section is compensated by the increase in the density of the PBHs near the galaxy center, so the capture rate is dominated by the spike region. Thus we recover the results in Table~\ref{tabl}, that the innermost region dominates binary capture rate, and the tidal effect is not prominent in affecting binary formation.

\begin{figure}
\centering
 \includegraphics[width=110mm]{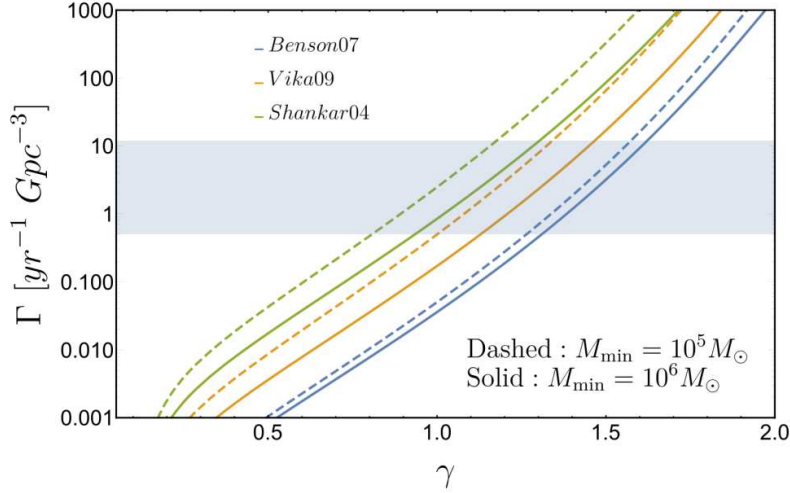}
 \caption{Binary formation rate density with 3 different halo mass functions, with solid and dotted lines showing different minimum halo masses, plotted against the power index of the initial PBH distribution, where for an NFW profile we look at $\gamma=1$. The shaded
region represents the rate $0.5 - 12 Gpc^{-3}yr^{-1}$ estimated by
LIGO (\citealt{2016PhRvX...6d1015A})}
\label{fig:ratedens}
\end{figure}

After the binary formation in a single MW-like galaxy has been found out, it can be convolved with the halo mass function to compute the binary formation rate per comoving volume and compared with LIGO's estimation of $0.5 - 12 \; Gpc^{-3}\,yr^{-1}$, which was later raised to $9.7-101 \; Gpc^{-3} yr^{-1}$ with all ten binary BH detections considered (\citealt{2016PhRvX...6d1015A, 2019PhRvX...9c1040A}). During this work there has been a similar study coming out that used the naive capture rate at the spike region and 3 SMBH mass functions to calculate the total rate density in halos hosting a SMBH (\citealt{2017arXiv170808449N}). The result is adopted here as Figure~\ref{fig:ratedens}. Note that the shaded region showing LIGO's event rate estimation should be shifted upward to reflect the most updated value. Since we have shown that the tidal effect does not change the binary formation rate by orders of magnitude, the conclusion drawn from this figure is unaltered. From the plot, even the most optimistic estimation can only reach the lower bound of LIGO's estimation near $\gamma=1$. Thus it is not very likely that halos with SMBHs alone gave rise to LIGO's detection.

Besides this sub-group, we also need to incorporate the contribution from halos without SMBHs, especially from those small halos whose signals actually dominate \citep{2017arXiv170808449N,2016PhRvL.116t1301B}. Thus we refer to \cite{2016PhRvL.116t1301B}, which calculated the total merger rate for all halos using the simple NFW profile. By including halos as small as $400 M_{\odot}$, the total merger rate is around $ 2 Gpc^{-3} yr^{-1}$. With these two results combined, we see that the total rate is still barely comparable to the lower limit of LIGO's estimation. Unless the actual physical situation deviates from the modeling noticeably, it is not likely that PBH DM with a monochromatic masss of $30 M_{\odot}$ alone explains for the GW signal.

\begin{figure}
\centering
 \includegraphics[width=100mm]{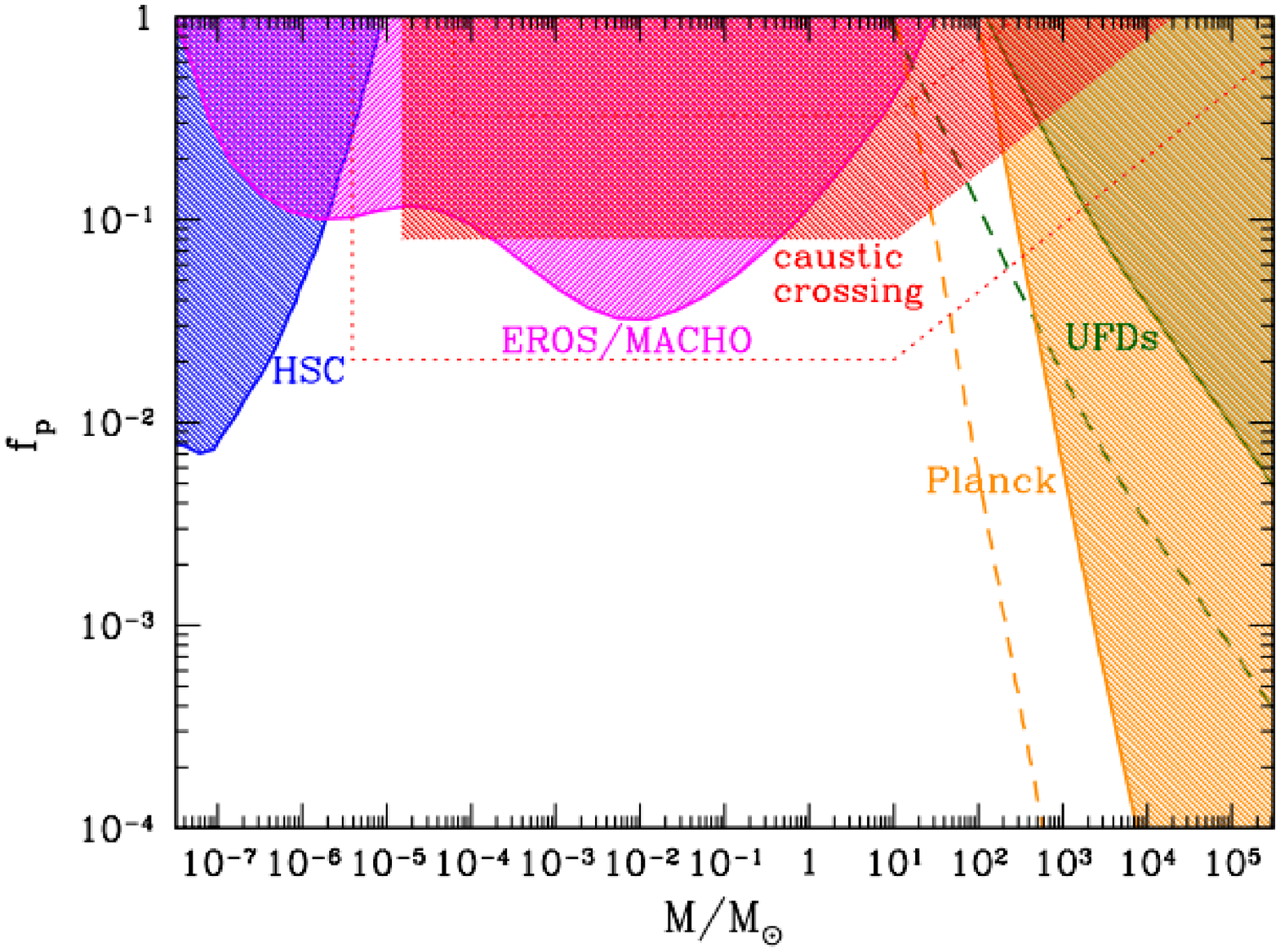}
 \caption{Constraints on the mass(M) and fraction of compact DM. Shaded regions show excluded range by caustic crossing of MACS J1149 LS1, M31 microlensing by Subaru/Hyper Suprime-Cam(HSC) (\citealt{2017arXiv170102151N}), EROS/MACHO microlensing (\citealt{2000ApJ...542..281A,2007A&A...469..387T}), ultra-faint dwarf galaxies (UFD) (\citealt{2016ApJ...824L..31B}) and Planck cosmic microwave background observations (Planck) (\citealt{2017PhRvD..95d3534A}). For caustic crossing, dotted lines show effect with different assumed transverse velocities, being twice and half of the fiducial value. For UFDs and Planck, dotted lines show more stringent limits compared with the conservative ones in solid lines.}
\label{fig:caucros}
\end{figure}

Even more importantly, a lensing event MACS J1149 LS1 was later observed, where a background star at z=1.49 experienced a transient magnification of several thousands and qualifies for a caustic crossing event (\citealt{2018NatAs...2..334K}). A caustic curve, or caustic, is the line joining the locations of the largest magnification on the source plane. Such geometrical property is influenced by the distribution of the lenses at the foreground. A follow-up study showed that the existence of abundant massive compact DM such as $30 M_{\odot}$ PBH would break the condition for the formation of the caustic. Monochromatic PBH DM of $30 M_{\odot}$ is thus further constrained on the mass-fraction plane as illustrated in Figure~\ref{fig:caucros} (\citealt{2018PhRvD..97b3518O}).    

We could see that the whole mass window of PBH on the order of $10 \sim 100 M_{\odot}$ has been well constrained to $\sim 10\%$ of total DM. As we have shown that the binary formation rate is proportional to the density squared as in equation (\ref{eqn:naiverate},\ref{eqn:effectiverate},\ref{eqn:survivalrate}), the new constraint from the caustic crossing event reduces the calculated binary formation rate in section \ref{sect:result} by 2 orders of magnitude. Thus it not only constrains monochromatic PBH to constitute a major fraction of DM, but implies that it is very unlikely that LIGO's detection signals originate from $30 M_{\odot}$ PBH DM. 

However, it was later argued that binary PBH could form via another channel in the early days of the universe, roughly at matter-radiation equality, that two formed PBHs close enough would have enough gravitational attraction to decouple from cosmic expansion (\citealt{2016PhRvL.117f1101S,2016PASJ...68...66H,2017PhRvD..96l3523A}). The corrseponding merger rate was found to coincide with LIGO's estimation if PBHs only constitute $\sim 1\%$ of DM. This is consistent with the current constraints on PBH DM mentioned before in Figure~\ref{fig:caucros}. Or rather, it further constrains PBH DM based on LIGO's detections. Another study took a step forward in computing the merger rate from early formed PBH binaries with two types of typical extended mass functions, namely power-law and log-normal, and found aggreement that LIGO constrains $M_{\odot}\sim 100 M_{\odot}$ PBHs to consitute $10^{-3}\sim 10^{-2}$ of DM (\citealt{2018ApJ...864...61C}). Thus we see that LIGO might indeed have detected mergers of binary PBHs formed in the early universe, yet they only account for around one percent of total DM at most. 

From Figure~\ref{fig:caucros} we see that PBH DM in the mass range across $10^{-7}\sim 10^6 M_{\odot}$ has been well constrained below $\sim 1$ percent. Constraints are also present across lower mass ranges, yet there still remains a major window around $10^{-15}\sim 10^{-11} M_{\odot}$ due to the debate over the constraints imposed by the capture of Neutron Stars (\citealt{2013PhRvD..87l3524C}), which relies on the premise that DM exists in the cores of Globular Clusters (\citealt{2018CQGra..35f3001S}). Thus it is still promising that PBHs constitute a major fraction of DM, but not of the masses typical of LIGO's detection.

\section{Conclusion}
\label{sect:concl}

LIGO's frequently detected binary BH mergers of $\sim 30 \; M_{\odot}$ can be fully attributed to PBH binaries formed in the early universe. The event rate constrains PBHs at $\sim 30 \; M_{\odot}$ to comprise DM below one persent. PBH DM could also form binaries via close-encounters in the late universe, which is dominated by the central region of the halo where DM desity is supposed to form a spike. At such vicinity of the SMBH at galaxy center, whose tidal effect is expected to be strong, we have shown that it has no significant impact on binary formation. When doing so, we have discovered an essential notion called 'loss zone' that complements the usage of 'loss cone' in evaluating the tidal effect on the fate of the object of interest. However, this late binary formation channel corresponds to a much lower event rate, approximately four orders smaller. Yet the loss zone notion is not restricted to this scenario only, and should apply wherever loss cone is used. 

All together, LIGO might have detected binary mergers of PBH DM formed in the early universe. Such detection also constrains the fraction of PBH around 30 solar mass in DM below one percent. Thus DM retains its mystery, and people should keep searching for alternative PBH windows and other candidate DM.

\begin{acknowledgements}
We would like to thank Dr. Ilias Cholis, Mr. Qinan Wang, Mr. Hong Tsun Wong, Mr. Jianju Tang and Mr. Renjiu Hu for the fruitful disccusions in carrying out this research work. 
\end{acknowledgements}

\bibliographystyle{raa}
\bibliography{references}

\end{document}